\documentclass[11pt]{article}
\usepackage{amssymb,amsmath,cite,graphicx,geometry,mathrsfs}
\catcode`\@=11

\@addtoreset{equation}{section}
\def\theequation{\arabic{section}.\arabic{equation}}
     
\catcode`\@=11
\def\thesection{\arabic{section}}

\def\appendix{\setcounter{section}{0}
        \def\thesection{Appendix \Alph{section}}
        \def\theequation{\Alph{section}.\arabic{equation}}}
\def\section{\@startsection{section}{1}{\z@}{3.5ex plus 1ex minus
   .2ex}{2.3ex plus .2ex}{\large\bf}}
   
\long\def\@makefntext#1{\parindent 0cm\noindent
\hbox to 1em{\hss$^{\@thefnmark}$}#1}

\newcommand{\captionfonts}{\small}
\makeatletter  
\long\def\@makecaption#1#2{%
  \vskip\abovecaptionskip
  \sbox\@tempboxa{{\captionfonts #1: #2}}%
  \ifdim \wd\@tempboxa >\hsize
    {\captionfonts #1: #2\par}
  \else
    \hbox to\hsize{\hfil\box\@tempboxa\hfil}%
  \fi
  \vskip\belowcaptionskip}
\makeatother   

\newcommand{\tq}{\triangleq}

\begin{document}
\begin{titlepage}
\vspace{.5in}
\begin{flushright}
February 2017\\  
\end{flushright}
\vspace{.5in}
\begin{center}
{\Large\bf
 Black Hole Entropy\\[1.1ex] from BMS Symmetry at the Horizon}\\  
\vspace{.4in}
{S.~C{\sc arlip}\footnote{\it email: carlip@physics.ucdavis.edu}\\
       {\small\it Department of Physics}\\
       {\small\it University of California}\\
       {\small\it Davis, CA 95616}\\{\small\it USA}}
\end{center}

\vspace{.5in}
\begin{center}
{\large\bf Abstract}
\end{center}
\begin{center}
\begin{minipage}{4.5in}
{\small 
Near the horizon, the obvious symmetries of a black hole spacetime---the 
horizon-preserving diffeomorphisms---are enhanced to a larger symmetry
group with a BMS${}_3$ algebra.  Using dimensional reduction and 
covariant phase space  techniques, I investigate this augmented symmetry, and 
show that it is strong enough to determine the black hole entropy. 
}
\end{minipage}
\end{center}
\end{titlepage}
\addtocounter{footnote}{-1}

\section{Introduction}

A striking feature of black hole thermodynamics is the universality of the 
Bekenstein-Hawking entropy.  Black holes, black strings, black rings, black 
branes, and black Saturns, in any dimension, with any charges and spins, with  
horizons arbitrary distorted by external fields, all have entropies given by the same 
simple expression,
\begin{align}
S_{\hbox{\tiny\it BH}} = \frac{A_{\hbox{\tiny\it hor}}}{4G\hbar}
\label{intro1}
\end{align}
where $A_{\hbox{\tiny\it hor}}$ is the horizon area.  Changing the action can change 
this formula, but only by another universal term.

The mystery deepens when one notes that many different models of the 
quantum black hole, from string theory to loop quantum gravity to induced gravity, 
can all yield the same entropy, even though they appear to count very different 
microstates \cite{Carlipz}.  Even in the elegant analysis of BPS black holes in 
string theory \cite{Vafa}, a separate computation is needed for each choice of 
dimension and each set of charges.  It seems clear that some underlying 
structure is yet to be found.

A natural guess for this deeper structure is that the degrees of freedom 
responsible for the entropy live on the horizon \cite{Bekenstein}.  But this is not 
enough: while it may explain the proportionality of entropy to area, there is no 
obvious reason for the coefficient of $1/4$ to be universal.  A more elaborate idea, 
first suggested (I believe) in \cite{Carlip1}, is that the entropy is governed by a 
horizon symmetry.  Two-dimensional conformal symmetry, in particular, has similar 
universal properties---the Cardy formula fixes the asymptotic behavior of the 
density of states in terms of a few parameters, independent of the details of the
theory \cite{Cardy}---and the possibility of a connection is appealing.  

This possibility was first confirmed for the (2+1)-dimensional BTZ black hole in 1998
\cite{Strominger,BSS}, and attempts to extend it to higher dimensions soon
followed \cite{Carlip2,Solodukhin}.  But while these efforts have had significant
successes---see \cite{Carlip3} for a review---they have been plagued by several  
serious limitations:
\begin{itemize}
\item The symmetries are almost always taken to be either at infinity or at a 
timelike ``stretched horizon'' (although with rare exceptions \cite{Paddy}).
While physics at infinity is very powerful, especially for asymptotically anti-de Sitter 
spaces, the symmetries by themselves cannot distinguish a black hole from, for 
instance, a star.  The stretched horizon more directly captures the properties of 
the black hole, but while the entropy has a well-defined limit at the horizon, 
other parameters typically blow up \cite{Dreyer,Koga} (again with occasional 
exceptions \cite{Dreyer2}).  Moreover, the definition of the stretched horizon 
is not unique, and different choices can lead to different entropies \cite{Silva,Carlip4}.
\item The standard canonical approach fails in what should be the simplest case,
two-dimensional dilaton gravity.  Symmetry generators are defined at  boundaries 
of spatial slices, and the zero-dimensional boundary of a one-dimensional slice 
is simply too small.  There are ad hoc fixes---lifting the theory to three dimensions 
\cite{Navarro} or artificially introducing an integral over time \cite{Mignemi}---but 
none is convincing.  
\item In higher dimensions, the relevant symmetries are those of the ``$r$--$t$ 
plane'' picked out by the horizon.  But to obtain a well-behaved symmetry algebra, 
one must introduce an extra ad hoc angular dependence of the parameters that
has no clear physical justification.
\end{itemize}

Here\footnote{ An expanded version of this work will appear in \cite{Carlipx}.} 
I show how to fix these problems.  The basic mistake, I argue, has been to try
to force the horizon symmetry into the form of a two-dimensional conformal
symmetry.  This was understandable: until recently, such a symmetry was the
only one known to be powerful enough to control the density of states.  But it 
has been recently discovered that a BMS${}_3$ (or Galilean conformal) symmetry 
has similar universal properties, including a generalized Cardy formula for the 
asymptotic density of states \cite{Bagchi}.

By using covariant phase space methods, introduced in this context in \cite{Carlip5} 
and elaborated in \cite{Barnich}, I show that the symmetry generators can be
expressed as integrals along the horizon \cite{Carlip6}, with no need for 
``stretching.''  I then demonstrate that a BMS${}_3$ symmetry appears in 
a completely natural way on the horizon, circumventing the problems of previous
efforts, and that it gives the correct counting of states.

\section{Dilaton gravity with null dyads}

The horizon $\Delta$ of a stationary black hole in any dimension has a preferred null 
direction, determined by the geodesics that generate the horizon.  A neighborhood  of 
$\Delta$ also has a preferred spatial coordinate, the proper distance from the horizon.
Together, these define a two-dimensional $r$--$t$ plane, in which most of the
interesting physics is expected to take place, since transverse derivatives are
red-shifted away near the horizon.  Hawking radiation, for instance, can be obtained 
by dimensional reduction to this plane \cite{Wilczek}.

Upon dimensional reduction and a field redefinition, the Einstein-Hilbert action 
becomes \cite{Kunstatter}
\begin{align}
I = \frac{1}{16\pi G}\int_M\!\left(\varphi R + V[\varphi]\right){\epsilon}
\label{a01}
\end{align}
where $\epsilon$ is the volume two-form. The scalar field $\varphi$, the dilaton, is 
the remnant of the transverse geometry, essentially the transverse area.  The resulting 
equations of motion are
\begin{subequations}
\begin{align}
&E_{ab} = \nabla_a\nabla_b\varphi - g_{ab}\Box\varphi + \frac{1}{2}g_{ab}V = 0 \label{ab5}\\
&R + \frac{dV}{d\varphi} = 0 \label{ab5b}
\end{align}
\end{subequations}
where the second equation follows from the divergence of the first.  
 
Let us choose a null dyad $(\ell_a,n_a)$, with $\ell^2= n^2 = 0$, normalized 
so that $\ell\cdot n = -1$.  For notational convenience, define $D=\ell^a\nabla_a$, 
${\bar D} = n^a\nabla_a$.  The metric and Levi-Civita tensor are then
\begin{align}
g_{ab} = - \left(\ell_an_b + n_a\ell_b\right) \qquad\qquad
\epsilon_{ab} =  \left(\ell_an_b - n_a\ell_b\right)
\label{a1}
\end{align}
The dyad is determined only up to a local Lorentz transformation,
$\ell^a \rightarrow e^\lambda\ell^a$, $n^a  \rightarrow e^{-\lambda} n^a$.
We can partially fix this freedom by choosing $n_a$ to have vanishing acceleration,
$n^b\nabla_bn^a = 0$; the remaining transformations are those for which 
$n^a\nabla_a\lambda=0$.  With this choice,  
\begin{alignat}{3}
&\nabla_a\ell_b = - \kappa n_a\ell_b \qquad\qquad
   && \nabla_a\ell^a = \kappa\nonumber\\
&\nabla_a n_b =  \kappa n_a n_b && \nabla_a n^a =0
\label{a2}
\end{alignat}
where $\kappa$ will be the surface gravity at a horizon.  Under variation of 
the dyad, (\ref{a2}) is preserved if
\begin{align}
&{\bar D}(\ell^c\delta n_c) = (D+\kappa)(n^c\delta n_c) \nonumber\\
&\delta\kappa = -  D(n^c\delta\ell_c) +  \kappa \ell^c\delta n_c 
   +  {\bar D}(\ell^c\delta\ell_c)
\label{a3a}
\end{align}

By considering the commutator $[\nabla_a,\nabla_b]\ell^b$, one may easily show that
\begin{align}
R =  2{\bar D}\kappa
\label{a4}
\end{align}
Below, I will also frequently use two identities:
\begin{align}
&[D,{\bar D}] = -\kappa{\bar D}   \label{ab4} \\
& df = -Df\,n_a - {\bar D}f\,\ell_a \quad\hbox{for any function $f$}  \label{abc1}
\end{align}
where in the latter I am treating $n_a$ and $\ell_a$ as one-forms.  Eqn.\
(\ref{abc1}) will be useful for integration by parts along the horizon.

\section{The covariant canonical formalism and symplectic structure \label{symp}}

The idea underlying the covariant canonical formalism is that for a theory with a unique
time evolution, the phase space, viewed as the space of initial data, can be
identified with the space of classical solutions \cite{Bombelli, Wald}.  This observation,
which can be traced back to Lagrange (see \cite{Bombelli}), means that we can formulate
all the usual ingredients of a Hamiltonian approach without ever having to break
general covariance by choosing a time slicing.

Consider a theory in an $n$-dimensional spacetime with fields $\Phi^A$ (for us,
$\varphi$ and $g$) and a Lagrangian density $L[\Phi]$, which we view as an 
$n$-form.  Under a general variation of the fields, $L[\Phi]$ changes as
\begin{align}
\delta L = E_A\delta\Phi^A + d\Theta[\Phi,\delta\Phi]
\label{a6}
\end{align}
where the equations of motion are $E_A=0$ and the last ``boundary'' term comes from 
integration by parts.  The symplectic current $\omega$ is defined as
\begin{align}
\omega[\Phi;\delta_1\Phi,\delta_2\Phi] 
   = \delta_1\Theta[\Phi,\delta_2\Phi] - \delta_2\Theta[\Phi,\delta_1\Phi]
\label{ab6}
\end{align}
and the symplectic form is
\begin{align}
\Omega[\Phi;\delta_1\Phi,\delta_2\Phi] 
    = \int_\Sigma \omega[\Phi;\delta_1\Phi,\delta_2\Phi]
    = \int_\Sigma \omega_{AB}\delta_1\Phi^A\wedge\delta_2\Phi^B
\label{ab7}
\end{align}
where $\Sigma$ is a Cauchy surface.  In keeping with the covariant phase space
philosophy, $\Omega[\Phi;\delta_1\Phi,\delta_2\Phi]$ depends on a classical solution
$\Phi$, which fixes a point in phase space, and is a two-form on the phase space.  The 
variations $\delta\Phi$ are thus tangent vectors to the space of classical solutions, that is, 
solutions of the linearized equations of motion.  The integral (\ref{ab7}) may depend 
on the choice of Cauchy surface, but only weakly: the symplectic current is a closed form, 
so integrals over two Cauchy surfaces $\Sigma_1$ and $\Sigma_2$ differ only by boundary 
terms that arise if  $\partial\Sigma_1\ne\partial\Sigma_2$.

As in ordinary mechanics, the symplectic form determines Poisson brackets 
and Hamiltonians.  In particular, given a family of transformations $\delta_\tau\Phi^A$
labeled by a parameter $\tau$, the Hamiltonian $H[\tau]$ is determined by the condition
\begin{align}
\delta H[\tau] = \Omega[\delta\Phi,\delta_\tau\Phi]
\label{ab8}
\end{align}
for an arbitrary variation $\delta\Phi$.  Indeed, this is just Hamilton's equation of motion,
\begin{align}
\delta_\tau\Phi^A = (\omega^{-1})^{AB}\frac{\delta H[\tau]}{\delta\Phi^B}
\label{ab9b}
\end{align}
The Poisson bracket of two Hamiltonians is then
\begin{align}
\left\{ H[\tau_1],H[\tau_2]\right\} = \Omega[\delta_{\tau_1}\Phi,\delta_{\tau_2}\Phi]
\label{ab10}
\end{align}

Specializing to dilaton gravity and using (\ref{a4}), it is straightforward 
to show that
\begin{align}
\Omega[(\varphi,g);\delta_1(\varphi,g),\delta_2(\varphi,g)] 
    = \frac{1}{8\pi G}\int_\Sigma \left[ \delta_1\varphi\,\delta_2(\kappa n_a) 
          + \delta_1({\bar D}\varphi)\delta_2\ell_a \right] - (1\leftrightarrow 2)
\label{a9}
\end{align}
where I am again treating $\ell_a$ and $n_a$ as one-forms on the (one-dimensional)
Cauchy surface $\Sigma$.
 
\section{Horizons and boundary conditions \label{bcs}}

For dilaton models obtained by dimensional reduction, $\varphi$ is essentially the
transverse area, and the natural definition of a local ``nonexpanding horizon'' 
$\Delta$---a null surface with vanishing expansion \cite{Ashtekar}---is that 
$D\varphi=0$ on $\Delta$.  This correctly determines the horizon from the purely 
two-dimensional viewpoint as well: on shell, $\Delta$ is a Killing horizon 
\cite{Kunstatter} and the boundary of a trapped region \cite{Cai}.  Exact black hole 
solutions in two dimensions have such horizons, with essentially the same Penrose
diagrams as those in higher dimensions \cite{Grumiller}.  

\begin{figure}
\centering
\begin{picture}(100,100)(0,-45)
\put(0,0){\line(1,1){50}}
\put(0,0){\line(1,-1){50}}
\put(100,0){\line(-1,1){50}}
\put(100,0){\line(-1,-1){50}}
\put(8,26){$\Delta$}
\put(79,26){$\mathscr{I}^+$}
\put(48,54){$i^+$}
\put(8,-31){$\Delta^{\!-}$}
\put(79,-31){$\mathscr{I}^-$}
\put(103,-2){$i^0$}
\put(-12,-2){$B$}
\put(0,0){\circle*{3}}
\end{picture}
\caption{Penrose diagram for the exterior of a black hole \label{fig1}}
\end{figure}
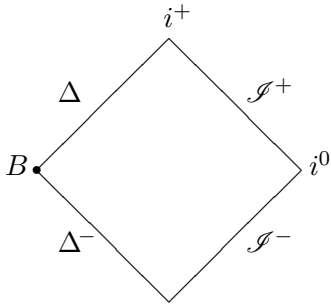
To study horizon symmetries in the covariant phase space formalism, we shall 
incorporate $\Delta$ as part of our Cauchy surface.  Let us focus on the exterior 
region of an asymptotically flat black hole, with the Penrose diagram of figure 
\ref{fig1}.  Take $\Sigma$ to be the union of the future horizon $\Delta$ and 
future null infinity $\mathscr{I}^+$, with ends at the bifurcation point $B$ and 
spacelike infinity.    The details of $\mathscr{I}^+$ are unimportant; the analysis 
below would be unchanged for asymptotically de Sitter or anti-de Sitter spaces.

Define  $\tq$ to mean ``equal on $\Delta$,'' where the horizon $\Delta$
is now determined by the requirement that $D\varphi\tq 0$.  We shall impose three 
``boundary conditions'' at this horizon:
\begin{enumerate}
\item $DR \tq 0$.
This is a requirement of stationary geometry on $\Delta$.  In higher dimensions, 
this condition follows automatically from the Raychaudhuri equation; here it must 
be imposed by hand, though it holds identically on shell.
\item The conformal class of the metric is fixed on $\Delta$, so $\ell^a\delta \ell_a \tq 0$
and $n^a\delta n_a \tq 0$.
This is in keeping with physical picture of conformal fluctuations of the metric 
as the relevant degrees of freedom.  I believe this condition can be relaxed, but at 
the cost of some complication.
\item The integration measure $n_a$ if fixed on $\Delta$.
In view of condition 2, this is the additional requirement that $\ell^a\delta n_a \tq 0$.
This is really a gauge-fixing condition, which can always be achieved by a suitable local 
Lorentz transformation.  Again, it may be possible to relax this requirement \cite{Carlipx}.
\end{enumerate}
These conditions simplify the symplectic form (\ref{a9}) considerably: for the portion 
lying on the horizon,
\begin{align}
\Omega_\Delta[(\varphi,g);\delta_1(\varphi,g),\delta_2(\varphi,g)] 
    = \frac{1}{8\pi G}\int_\Delta \left[ \delta_1\varphi\,\delta_2\kappa 
    - \delta_1\varphi\,\delta_2\kappa \right] n_a
\label{ac7}
\end{align}
One  subtlety remains, though.  A variation of $\varphi$ will
typically ``move the horizon,'' changing the locus of points $D\varphi=0$.  
This will not matter for the symplectic form, since $\Omega_\Delta$ is independent
of the integration contour.  More precisely, if $\delta_\zeta$ is a transverse diffeomorphism 
generated by a vector field $\zeta^a = {\bar\zeta} n^a$,  
\begin{align}
\Omega_\Delta[(\varphi,g);\delta(\varphi,g),\delta_\zeta(\varphi,g)] 
    = -\frac{1}{8\pi G}\int_\Delta {\bar\zeta}(D+\kappa){\bar D}\varphi \,\delta n_a
    + \frac{1}{8\pi G}({\bar\zeta}{\bar D}\varphi)\Bigl|_{\partial\Delta} 
\label{ac10}
\end{align}
The bulk term vanishes by virtue of the boundary condition $\delta n_a=0$,
and the boundary term will vanish provided that ${\bar\eta}=0$ at $\partial\Delta$.

For the variation of an object such as a Hamiltonian defined as an integral over $\Delta$, 
however, we shall have to take this change into account.  The diffeomorphism 
needed to ``move the horizon back'' is determined by the condition that
\begin{align}
\delta(D\varphi) + \zeta^a\nabla_a(D\varphi) \tq 0 \ \Rightarrow \ 
\zeta^a = {\bar\zeta}n^a = -\frac{D\delta\varphi}{{\bar D}D\varphi}n^a
\label{ac8}
\end{align}
and hence
\begin{align}
\delta \int_\Delta \mathscr{H}\,n_a 
   = \int_\Delta( \delta\mathscr{H} + \zeta^a\nabla_a \mathscr{H})n_a
\label{ac9}
\end{align}

\section{Symmetries and approximate symmetries \label{symm}}

The action (\ref{a01}) is, of course, invariant under diffeomorphisms, including   
horizon ``supertranslations'' \cite{Donnay} generated by vector fields $\xi^a = \xi\ell^a$.  
By condition 3 of the preceding section, we must supplement such diffeomorphisms by 
local Lorentz transformations $\delta\lambda = D\xi$ to ensure that $\ell^a\delta_\xi n_a=0$.  
By (\ref{a3a}), this requires that ${\bar D}\xi\tq0$.   We thus have an invariance
\begin{align}
&\delta_\xi\ell^a = 0, \quad \delta_\xi n^a = -(D+\kappa)\xi \,n^a \nonumber\\
&\delta_\xi g_{ab} = -(D+\kappa)\xi \, g_{ab} \nonumber\\
&\delta_\xi \varphi = \xi D\varphi \quad \hbox{with ${\bar D}\xi\tq0$}
\label{c1}
\end{align}
 
As noted long ago, though \cite{Carlip6}, the action also has an \emph{approximate} 
invariance under a certain shift of the dilaton near a black hole horizon, with an
approximation that can be made arbitrarily good by restricting the transformation  
to a small enough neighborhood of $\Delta$.    Consider a variation 
\begin{align}
{\hat\delta}_\eta\varphi = \nabla_a(\eta\ell^a) = (D+\kappa)\eta  \quad \hbox{with ${\bar D}\eta\tq0$}
\label{c1a}
\end{align}
(where the hat in $\hat\delta$ distinguishes it from a supertranslation).
The action transforms as 
\begin{align}
{\hat\delta} _\eta I &= \frac{1}{16\pi G}\int_M\!  \left(R + \frac{dV}{d\varphi}\right)
  {\hat\delta}_\eta\varphi\, {\epsilon}  
 = -\frac{1}{16\pi G}\int_M \eta\left[ DR + \frac{d^2V}{d\varphi^2}D\varphi  
   \right]{\epsilon}
\label{c2}
\end{align}
But $D\varphi$ and $DR$ both vanish at the horizon, so  the variation (\ref{c2}) 
can be made as small as one wishes by choosing $\eta$ to fall off fast enough away 
from $\Delta$.   

This is not quite enough: while the transformation (\ref{c1a}) does not directly act on 
the curvature, the change of $\varphi$ moves the horizon, and $DR$ may no longer 
vanish at the new location.  The displacement of the horizon is characterized by the 
diffeomorphism (\ref{ac8}), and can be compensated  with a ``small''  (order $D\varphi$) 
Weyl transformation of the metric to restore the condition $DR\tq0$:
\begin{align}
{\hat\delta}_\eta g_{ab} = {\hat\delta}\omega_\eta\,g_{ab} \quad \hbox{with}\ \ 
   {\hat\delta}\omega_\eta = X\frac{D\varphi}{{\bar D}D\varphi} , \ \ 
   {\bar\zeta}{\bar D}DR + 2D(D+\kappa)X \tq 0
\label{c7}
\end{align}
On shell, a short calculation gives an explicit expression for $X$:
\begin{align}
{\bar\zeta}{\bar D}DR + 2D(D+\kappa)X &\tq
   \frac{d^2V}{d\varphi^2}D(D+\kappa)\eta + 2D(D+\kappa)X \tq 0 \ \Rightarrow \
   X \tq -\frac{1}{2}\frac{d^2V}{d\varphi^2}\eta
\label{cx2}
\end{align}
Like (\ref{c1a}), the Weyl transformation (\ref{c7})  changes the action only by terms 
proportional to $D\varphi$, which can be made arbitrarily small by choosing $\eta$ to fall off
fast enough away from $\Delta$.  

We must also check the variation of the equations of motion (\ref{ab5})--(\ref{ab5b}).
These are, of course, preserved by diffeomorphisms, so we need only consider the
transformations (\ref{c1a}) and (\ref{c7}).  Since we are assuming that $\eta$ falls off rapidly 
away from the horizon, it is enough to check the variations at $\Delta$.  Note that 
after varying an equation of motion, we can put the system on shell---a variation of an 
equation of motion by s symmetry need only vanish up to equations of motion.

A straightforward computation then shows that on shell,
\begin{subequations}
\begin{align}
g^{ab}{\hat\delta}_\eta E_{ab} \tq 2(D+\kappa){\bar D}{\hat\delta}_\eta\varphi 
   +  \frac{dV}{d\varphi} {\hat\delta}_\eta\varphi &\tq 0 \label{ce2}\\
n^an^b{\hat\delta}_\eta E_{ab} \tq {\bar D}^2{\hat\delta}_\eta\varphi
  - {\bar D}\varphi{\bar D}{\hat\delta}_\eta\omega &\tq 0 \label{ce1}\\
{\hat\delta}_\eta\left( R + \frac{dV}{d\varphi}\right) 
   \tq {\hat\delta}_\eta R + \frac{d^2V}{d\varphi^2}{\hat\delta}_\eta\varphi &\tq 0
   \label{ce4}
\end{align}
\end{subequations}
This leaves the variation $\ell^a\ell^b{\hat\delta}_\eta E_{ab}$, which is not zero, but
instead matches the anomalous variation of the stress-energy tensor in a conformal 
field theory.  Indeed, if we set $E_{ab} = 8\pi G T_{ab}$,  
we find
\begin{align}
\ell^a\ell^b{\hat\delta}_\eta T_{ab} \tq \frac{1}{8\pi G} (D-\kappa)D(D+\kappa)\eta
\label{ca6}
\end{align}
which is just the anomaly for a conformal field theory with a central charge
proportional to $1/G$.  One might worry that the anomaly could spoil the
covariant phase space construction of section \ref{symp}, since the closure of the
symplectic current (\ref{ab6}) relies on the classical field equations.  Fortunately, this is
not a problem: the only dangerous term is proportional to $\eta \,n^a\delta n_a$, which 
is zero at the horizon because of our boundary conditions and falls off like $\eta$ 
away from the horizon.

\section{Canonical generators and their algebra}

At the horizon, the two symmetries of the preceding section obey an algebra
\begin{alignat}{3}
&[\delta_{\xi_1}, \delta_{\xi_2}] f \tq \delta_{\xi_{12}}f \qquad
  && \hbox{with}\ \ \xi_{12} = - (\xi_1D\xi_2 - \xi_2D\xi_1)  \nonumber\\
&[{\hat\delta}_{\eta_1}, {\hat\delta}_{\eta_2}] f \tq 0 \nonumber\\
&[\delta_{\xi_1}, {\hat\delta}_{\eta_2}] f \tq {\hat\delta}_{\eta_{12}}f \qquad
  && \hbox{with}\ \ \eta_{12} = - (\xi_1D\eta_2 - \eta_2D\xi_1) 
\label{d1}
\end{alignat}
This may be recognized as a BMS$_3$ algebra, or equivalently a Galilean conformal algebra 
\cite{GCA}.  We must now ask whether these transformations can be realized 
canonically as in (\ref{ab8}), that is, whether there exist generators that satisfy
\begin{subequations}
\begin{align}
\delta L[\xi] &= \frac{1}{8\pi G}\int_\Delta\left[ \delta\varphi\,\delta_\xi\kappa 
    - \delta_\xi\varphi\,\delta\kappa \right] n_a
    = \frac{1}{8\pi G}\int_\Delta\left[ \delta\varphi\,D(D+\kappa)\xi 
    - \xi D\varphi\,\delta\kappa \right] n_a  \label{d2} \\
\delta M[\eta] &= \frac{1}{8\pi G}\int_\Delta\left[ \delta\varphi\,{\hat\delta}_\eta\kappa 
    - {\hat\delta_\eta}\varphi\,\delta\kappa \right] n_a  
    = \frac{1}{8\pi G}\int_\Delta\left[ -\delta\kappa(D+\kappa)\eta
    + \frac{1}{2}\frac{D\delta\varphi}{{\bar D}D\varphi}\eta\frac{d^2V}{d\varphi^2}D\varphi
    \right]
\label{d6}
\end{align}
\end{subequations}
where variations of the generators must include the horizon displacement described 
by (\ref{ac9}), and the covariant phase space formalism allows us to impose the equations 
of motion after variation. Such generators exist:
\begin{subequations}
\begin{align}
L[\xi] &= \frac{1}{8\pi G}\int_\Delta\left[\xi D^2\varphi - \kappa\xi D\varphi\right]n_a
\label{d3} \\
M[\eta] &= \frac{1}{8\pi G}\int_\Delta \eta
\left(D\kappa - \frac{1}{2}\kappa^2\right)n_a
\label{d7}
\end{align}
\end{subequations}
Using (\ref{ab10}), we find Poisson brackets\footnote{The first of these holds 
even if $D\varphi\ne0$.  The second and third do not---the $\eta$ transformations are
symmetries only on a horizon---but the deviations are of order $(D\varphi)^2$.}
\begin{subequations}
\begin{align}
&\left\{L[\xi_1],L[\xi_2]\right\} = L[\xi_{12}] \label{d11a}\\
&\left\{M[\eta_1],M[\eta_2]\right\} \tq 0 \label{d11b}\\
&\left\{L[\xi_1],M[\eta_2]\right\} \tq M[\eta_{12}] 
   + \frac{1}{16\pi G}\int_\Delta \left(D\xi_1 D^2\eta_2 - D\eta_2 D^2\xi_1\right)n_a
\label{d11c}
\end{align}
\end{subequations}
with $\xi_{12}$ and $\eta_{12}$ as in (\ref{d1}).   
The canonical generators thus give a representation of the symmetry algebra, but with an 
added off-diagonal central term.

\section{Modes, zero-modes, and entropy}

It is well known that for a unitary theory with a two-dimensional conformal symmetry,
the asymptotic density of states---the entropy---is determined, via the Cardy
formula, by the central charge \cite{Cardy}.  The same is true for a theory with a 
BMS$_3$ algebra \cite{Bagchi}: with a mode decomposition  
\begin{align}
&i\left\{L_m,L_n\right\} = (m-n)L_{m+n} \nonumber\\
&i\left\{M_m,M_n\right\} = 0 \nonumber\\
&i\left\{L_m,M_n\right\} = M_{m+n} + c_{\scriptscriptstyle LM}m(m^2-1)\delta_{m+n,0} 
\label{e1}
\end{align}
the asymptotic behavior of the entropy is
\begin{align} 
S \sim 2\pi h_{\scriptscriptstyle L}\sqrt{\frac{c_{\scriptscriptstyle LM}}{2h_{\scriptscriptstyle M}}}
\label{e2}
\end{align}
where $h_{\scriptscriptstyle L}$ and $h_{\scriptscriptstyle M}$ are the eigenvalues of $L_0$ and $M_0$.

To use this result, we first need a mode decomposition.  For a black hole with constant surface gravity, 
the relevant modes take the form $e^{in\kappa v}$, where $v$ is the advanced time along the horizon, 
normalized so that $\ell^a\nabla_av=1$.  We can generalize this by defining a phase $\psi$ such
that
\begin{align}
D\psi \tq \kappa, \quad {\bar D}\psi \tq 0 \quad\Rightarrow\ \ d\psi\tq -\kappa n_a
\label{e3}
\end{align}
The modes are then
\begin{align}
\zeta_n \tq \frac{1}{\kappa}e^{in\psi} \qquad\hbox{(where $\zeta$ is either $\xi$ or $\eta$)}
\label{e4}
\end{align}
with a prefactor chosen so
$\{\zeta_m,\zeta_n\} = \zeta_mD\zeta_n - \zeta_nD\zeta_m = -i(m-n)\zeta_{m+n}$.
With this moding,  
\begin{align}
\frac{1}{16\pi G}\int_\Delta \left(D\xi_m D^2\eta_n - D\eta_n D^2\xi_m\right)n_a
   = \frac{i}{8\pi G}\int_\Delta mn^2e^{i(m+n)\psi}d\psi
\label{e6}
\end{align}
If we take the integral to be over a single period---essentially mapping the problem
to a circle, as is standard in conformal field theory---we obtain a central charge in (\ref{d11c})
of
\begin{align}
c_{\scriptscriptstyle LM} = \frac{1}{4G}
\label{e7}
\end{align} 

We also need the zero-modes of  $L$ and $M$.  For $M$, this is straightforward: from (\ref{d7}),
\begin{align}
h_{\scriptscriptstyle M} = M[\eta_0] = -\frac{1}{16\pi G}\int_\Delta \kappa n_a 
   = \frac{1}{16\pi G}\int d\psi = \frac{1}{8G}
\label{e8}
\end{align}
For $L$, the ``bulk'' contribution to $L[\xi_0]$ vanishes.  But $L$, unlike $M$, 
has a boundary contribution.  Indeed,  the variation leading to (\ref{d2}) involves integration 
by parts, with a boundary term
\begin{align}
\delta L[\xi] = \dots + \frac{1}{8\pi G}\left(\xi D\delta\varphi 
   - (D+\kappa)\xi\,\delta\varphi\right)\Bigl|_{\partial\Delta}
\label{e9}
\end{align}
From (\ref{ac10})--(\ref{ac8}), we must set $D\delta\varphi$ to zero at  $\partial\Delta$,  
but we should certainly not hold $\varphi$ itself fixed, since that would fix $\varphi$ along the 
entire horizon, eliminating the $\eta$ symmetry.  Instead,  we should fix the conjugate
variable $\kappa$ at $\partial\Delta$.  This gives a boundary contribution at the bifurcation 
point of
\begin{align}
h_{\scriptscriptstyle L} = \frac{1}{8\pi G} \varphi (D + \kappa)\xi_0 \,\Bigl|_{\partial\Delta} 
      = \frac{\varphi_+}{8\pi G}
\label{e10}
\end{align}
where $\varphi_+$ is the value of $\varphi$ at the bifurcation point $B$ of figure \ref{fig1}.
Inserting (\ref{e7}), (\ref{e8}), and (\ref{e10}) into (\ref{e2}), we finally obtain  
\begin{align}
S = \frac{\varphi_+}{4G}
\label{e11}
\end{align}
which is precisely the correct Bekenstein-Hawking entropy.
 
\section{Conclusions}

We have seen that black hole entropy is indeed governed by horizon symmetries.
In contrast to previous attempts, this derivation requires no stretched horizon and 
no extra angular dependence or other ad hoc ingredients.  The main assumptions are 
merely that dimensional reduction is possible and that the horizon obeys the ``boundary 
conditions'' of section \ref{bcs}.

How should we think about the resulting BMS symmetry?  It is not a gauge symmetry:
our counting arguments imply that states are not invariant, but transform under
high-dimensional representations.  Nor is it quite a standard asymptotic symmetry:
while we can view the horizon as a sort of boundary, it is a boundary that exists
only for a restricted class of field configurations.  Physically, we are asking a question
of conditional probability---\emph{if} a black hole is present, what are its
properties?---and the symmetries reflect this condition.

There are obvious directions for generalization.  Dimensional
reduction focuses our attention on the relevant parts of the geometry, but it would
be good to explicitly lift the argument to higher dimensions.  We should clarify the 
relationship between the symmetries of this paper and other appearances of BMS 
symmetry at the horizon \cite{Donnay,Eling,Hawking,Fareghbal,Afshar}, as well 
as the related horizon symmetry used by Wall to prove the generalized second law 
\cite{Wall}.  It should be feasible to significantly relax the boundary conditions of 
section \ref{bcs}.   It  may also be possible to make the concept of ``approximate 
symmetry'' in section \ref{symm} more precise.  In this regard, recall that the 
shift parameter $\eta$ appears in the variation of the action with no transverse 
derivatives, and can also be rescaled by a constant without changing the algebra, 
so both its value and its support can be made arbitrarily small.

Finally, if this symmetry is really responsible for the universal properties of black hole entropy,
one might expect to find it hidden in other derivations of entropy.  Preliminary steps
in this direction have been taken for loop quantum gravity \cite{Carlip7}, for induced
gravity \cite{Frolov}, and perhaps for near-extremal black holes in string theory 
\cite{Carlip8}, but none of these attempts has exploited the full BMS symmetry.  Ideally, we  
could hope to do more: perhaps this symmetry can be used to couple the black hole to matter
and obtain Hawking radiation, as Emparan and Sachs did for the (2+1)-dimensional
black hole \cite{Emparan}.
\vspace{1.5ex}
\begin{flushleft}
\large\bf Acknowledgments
\end{flushleft}

This research was supported by the US Department of Energy under grant DE-FG02-91ER40674.


\begin{thebibliography}{99}
\bibitem{Carlipz} S.\ Carlip, in \emph{Quantum Mechanics of Fundamental Systems: 
   the Quest for Beauty and Simplicity}, edited by M.\ Henneaux and J.\ Zanelli (Springer, 
   New York, 2009),  arXiv:0807.4192.
\bibitem{Vafa} A.\ Strominger and C.\ Vafa, Phys.\ Lett.\ B379 (1996) 99,
   arXiv:hep-th/9601029.
\bibitem{Bekenstein}  J.~D.\ Bekenstein, Phys.\ Rev.\ D7  (1973) 2333.
\bibitem{Carlip1} S.\ Carlip, in \emph{Field Theory, Integrable Systems 
   and Symmetries}, edited by F.\ Khanna and L.\ Vinet  (Les Publications 
   CRM, Montreal, 1997), arXiv:gr-qc/9509024.
\bibitem{Cardy} H.~W.~J.\ Bl{\"o}te, J.~A.\ Cardy, and M.~P.\ Nightingale, 
   Phys.\ Rev.\ Lett.\ 56 (1986) 742.
\bibitem{Strominger} A.\ Strominger,  JHEP 9802 (1998) 009, 
   arXiv:hep-th/9712251.
\bibitem{BSS} D.\ Birmingham, I.\ Sachs, and S.\ Sen,  Phys.\ Lett.\ B424 (1998)
   27, arXiv:hep-th/9801019.
\bibitem{Carlip2} S.\ Carlip,  Phys.\ Rev.\ Lett.\ 82 (1999) 2828, 
   arXiv:hep-th/9812013.
\bibitem{Solodukhin} S.~N.\ Solodukhin, Phys.\ Lett.\ B454, 213 (1999),
    arXiv:hep-th/9812056.
\bibitem{Carlip3} S.\ Carlip, Entropy 13 (2011) 1355, arXiv:1107.2678.
\bibitem{Paddy} B.~R.\ Majhi and T.\ Padmanabhan, Phys.\ Rev.\ D86 (2012) 
   101501, arXiv:1204.1422.
\bibitem{Dreyer} O.\ Dreyer, A.\ Ghosh, and J.\ Wisniewski, Class.\ Quant.\ Grav.\
   18 (2001) 1929, arXiv:hep-th/0101117.
\bibitem{Koga} J.\ Koga, Phys.\ Rev.\  D64 (2001) 124012, arXiv:gr-qc/0107096.
\bibitem{Dreyer2} O.\ Dreyer, A.\ Ghosh, and A.\ Ghosh. Phys.\ Rev.\ D89 
   (2014) no.2, 024035, arXiv:1306.5063.
\bibitem{Silva} S.\ Silva, Class.\ Quant.\ Grav.\ 19 (2002) 3947,
   arXiv:hep-th/0204179.
\bibitem{Carlip4} S.\ Carlip,  JHEP 1104 (2011), 076, arXiv:1101.5136;  
   Erratum-ibid.\ 1201 (2012) 008.
\bibitem{Navarro} J.~M.\ Izquierdo, J.\ Navarro-Salas, and P.\ Navarro,
   Class.\ Quant.\ Grav.\ 19 (2002) 563, arXiv:hep-th/0107132.
\bibitem{Mignemi} M.\ Cadoni and S.\ Mignemi, Phys.\ Rev.\ D59 (1999)
   081501, arXiv:hep-th/9810251.
\bibitem{Bagchi} A.\ Bagchi, S.\ Detournay, R.\ Fareghbal, and J.\ Sim{\'o}n,
   Phys.\ Rev.\ Lett.\ 110 (2013) 141302, arXiv:1208.4372.
\bibitem{Carlip5} S.\ Carlip,  Class.\ Quant.\ Grav.\ 16 (1999) 3327, 
   arXiv:gr-qc/9906126.
\bibitem{Barnich} G.\ Barnich and F.\ Brandt, Nucl.\ Phys.\ B633 (2002) 3,
   arXiv:hep-th/0111246.
\bibitem{Carlip6} S.\ Carlip, Phys.\ Rev.\ Lett.\ 88 (2002) 241301,
   arXiv:gr-qc/0203001.
\bibitem{Carlipx} S.\ Carlip, in preparation.
\bibitem{Wilczek} S.~P.\ Robinson and F.\ Wilczek, Phys.\ Rev.\ Lett.\ 95 (2005)
   011303, arXiv:gr-qc/0502074.
\bibitem{Kunstatter} D.\ Louis-Martinez and G.\ Kunstatter, Phys.\ Rev.\ D52 (1995) 
   3494, arXiv:gr-qc/9503016.
\bibitem{Bombelli} A.\ Ashtekar, L.\ Bombelli, and O.\ Reula, in \emph{Analysis, 
   Geometry and Mechanics: 200 Years After Lagrange}, edited by M.\ Francaviglia
   (North-Holland, Amsterdam, 1991).
\bibitem{Wald} J.\ Lee and R.~M.\ Wald, J.\ Math.\ Phys.\ 31 (1990) 725.
\bibitem{Ashtekar} A.\ Ashtekar, C.\ Beetle, and S.\ Fairhurst, Class.\ Quant.\ Grav.\
   16 (1999) L1, arXiv:gr-qc/9812065.
\bibitem{Cai} R.-G.\ Cai and L.-M.\ Cao, arXiv:1609.08306.
\bibitem{Grumiller} D.\ Grumiller, W.\ Kummer, and D.~V.\ Vassilevich, Phys.\ Rept.\
   369 (2002) 327, arXiv:hep-th/0204253.
\bibitem{Donnay} L.\ Donnay, G.\ Giribet, H.~A.\ Gonz{\'a}lez, and M.\  Pino,
    Phys.\ Rev.\ Lett.\ 116 (2016)  091101, arXiv:1511.08687.
\bibitem{GCA} A.\ Bagchi, Phys.\ Rev.\ Lett.\ 105 (2010) 171601, arXiv:1006.3354.
\bibitem{Eling} C.\ Eling, JHEP 1607 (2016) 065, arXiv:1605.00183.
\bibitem{Hawking} S.~W.\ Hawking, M.~J.\ Perry, and A.\ Strominger, arXiv:1611.09175.
\bibitem{Fareghbal} R.\ Fareghbal and A.\ Naseh, JHEP 1406 (2014) 134, 
    arXiv:1404.3937.
\bibitem{Afshar} H.\ Afshar, D.\ Grumiller, W.\ Merbis, A.\ Perez, D.\ Tempo, 
    and R.\ Troncoso, arXiv:1611.09783.
\bibitem{Wall} A.~C.\ Wall, Phys.\ Rev.\ D 85 (2012) 104049, arXiv:1105.3445.
\bibitem{Carlip7} S.\ Carlip, Phys.\ Rev.\ Lett.\ 115 (2015) 071302, arXiv:1503.02981.
\bibitem{Frolov} V.~P.\ Frolov, D.\ Fursaev, and A.\ Zelnikov, JHEP 0303 (2003) 038,
   arXiv:hep-th/0302207.
\bibitem{Carlip8} S.\ Carlip, Phys.\ Rev.\ Lett.\ 99 (2007) 021301, arXiv:gr-qc/0702107.
\bibitem{Emparan}  R.\ Emparan and I.\ Sachs, Phys.\ Rev.\ Lett.\ 81 (1998) 2408,
   arXiv:hep-th/9806122.
\end{thebibliography}
\end{document}